\documentclass[12pt]{article}%
\usepackage{amsmath}
\usepackage{graphicx}%
\usepackage{amsfonts}%
\usepackage{amssymb}
\begin{document}

\title{Correlation functions, Bell's inequalities and the fundamental conservation laws}
\author{C. S. Unnikrishnan\thanks{E-mail address: unni@tifr.res.in}\\\textit{Gravitation Group, Tata Institute of Fundamental Research, }\\\textit{Homi Bhabha Road, Mumbai - 400 005, India}}
\date{}
\maketitle

\begin{abstract}
I derive the correlation function for a general theory of two-valued spin
variables that satisfy the fundamental conservation law of angular momentum.
The unique theory-independent correlation function is identical to the quantum
mechanical correlation function. I prove that any theory of correlations of
such discrete variables satisfying the fundamental conservation law of angular
momentum violates the Bell's inequalities. Taken together with the Bell's
theorem, this result has far reaching implications. No theory satisfying
Einstein locality, reality in the EPR-Bell sense, and the validity of the
conservation law can be constructed. Therefore, all local hidden variable
theories are incompatible with fundamental symmetries and conservation laws.
Bell's inequalities can be obeyed only by violating a conservation law. The
implications for experiments on Bell's inequalities are obvious. The result
provides new insight regarding entanglement, and its measures.

\end{abstract}

\medskip

 \bigskip

The main result of this paper is the proof that a general physical theory of
correlations of two-valued spin projections violate the Bell's inequalities if
the theory satisfies the conservation law for angular momentum on the average.
The proof is applicable to local hidden variable theories, as well as to other
theories like the non-local hidden variable theories, and to quantum mechanics
itself, with the only assumption that they respect the conservation law for
the ensemble average. The proof follows from the result that any such theory
predicts a unique correlation function that is in fact identical to the
quantum mechanical correlation function. Since such fundamental conservation
laws have their origin in space-time symmetries, and since we expect that the
only viable theories of the physical world are those satisfying these
conservation laws, this result has far reaching significance. The physical
situation we consider is the measurements of spin-projections of spin-half
particles or photons with their total angular momentum zero. The proof is
applicable to other values of total angular momentum.

A local hidden variable theory is a classical statistical theory meant to
replace quantum mechanics. Such theories were proposed with the hope that one
could preserve classical notions like locality, and reality of events and
history in space and time, and yet reproduce the statistical results of
quantum mechanics \cite{belin,bell}. The randomness in the measured results of
an observable in such theories is mapped to the random values taken by certain
hidden or unobservable classical variables. John Bell's analysis of local
hidden variable theories resulted in the celebrated Bell's inequalities
\cite{bell}. These represent an upper limit on the correlation expected in
such theories between results of measurements on separated correlated quantum
systems. In the standard formulations, the correlation and its upper limit in
a local hidden variable theory are typically smaller than what is predicted in
the quantum mechanical description, for a wide range of settings of the
measurement apparatus. Thus, quantum mechanical correlations violate the
Bell's inequalities.

Several experiments have been done in the past, and several are in planning
and execution to test the Bell's inequalities, and to thereby test the
viability of local hidden variable theories \cite{exp}. Essentially all
experiments to date find that the Bell's inequalities are violated, and that
the measured correlations are in fact larger than the upper limit specified by
the Bell's inequalities. The experiments also confirm with high precision that
the quantum mechanical prediction for the correlation is what is favoured.

Bell's derivation of the inequalities assumed Einstein locality, and local
hidden variables that could determine outcomes of measurements, and he dealt
with two-valued observables. The physical setting is that of two observers $A$
and $B$ making measurements on two correlated particles individually at
space-like separated regions with time information such that these results can
be correlated event by event.

Locality in hidden variable theories is represented by functional restrictions
on the outcomes $A$ and $B$ of measurements at the two locations.%

\begin{equation}
A(\mathbf{a,h})=\pm1,\quad B(\mathbf{b,h})=\pm1
\end{equation}

$A$ and $B$ denote the outcomes $+1$ or $-1$ of measurements $A$ and $B,$ and
$\mathbf{a}$ and $\mathbf{b}$ denote the settings of the analyzer or the
measurement apparatus for the first particle and the second particle
respectively. $\mathbf{h}$ are hidden variables associated with the outcomes.
The Bell correlation function \cite{bell} is of the form
\begin{equation}
P(\mathbf{a},\mathbf{b})=\int d\mathbf{h}\rho(\mathbf{h})A(\mathbf{a,h}%
)B(\mathbf{b,h}),\mathrm{{where}\int d\mathbf{h}\rho(\mathbf{h})=1}%
\end{equation}
This is an average over the product of the measurement results. The
experimenter calculates the observed correlation using the formula
\begin{equation}
P(\mathbf{a},\mathbf{b})=\frac{1}{N}\sum(A_{i}B_{i})
\end{equation}
from the observed outcomes $A_{i}$ and $B_{i}.$ The essence of Bell's theorem
is that the function $P(\mathbf{a},\mathbf{b})$ has distinctly different
dependences on the relative angle between the analyzers for a local hidden
variable description and for quantum mechanics. If $\mathbf{a}$ and
$\mathbf{a}^{\prime}$ are two settings of the apparatus at $A,$ and
$\mathbf{b},\mathbf{b}^{\prime}$ are settings at $B,$ then it was shown
\cite{bell} that the combination of correlations \
\begin{equation}
M_{LHVT}=\left|  P(\mathbf{a},\mathbf{b})-P(\mathbf{a},\mathbf{b}^{\prime
})\right|  +\left|  P(\mathbf{a}^{\prime},\mathbf{b}^{\prime})+P(\mathbf{a}%
^{\prime},\mathbf{b})\right|  \leq2
\end{equation}

The quantum mechanical correlation for the same experiment is given by
\begin{equation}
P(\mathbf{a},\mathbf{b})_{QM}=-\mathbf{a}\cdot\mathbf{b}%
\end{equation}
This is the expectation value of the operator $\sigma_{1}\cdot\mathbf{a\otimes
\sigma}_{2}\cdot\mathbf{b}$ for the singlet state (When expressed in units of
$\hbar/2$. In terms of values of the projections of spin, the correlation
function is $P(\mathbf{a},\mathbf{b})_{QM}=-\mathbf{a}\cdot\mathbf{b~\hbar
}^{2}/4$). There are several combinations of the set $\left\{  \mathbf{a}%
,\mathbf{a}^{\prime},\mathbf{b},\mathbf{b}^{\prime}\right\}  $ for which
$M_{QM}$ exceeds $M_{LHVT}$ of local hidden variable theories. Thus quantum
mechanical correlations violates Bell's inequalities.

In what follows, we start with the assumption the validity of the fundamental
conservations laws, like the conservation of angular momentum, on the average
(for probabilities) and calculate the general correlation function that is
allowed for two-valued observables. We make no assumptions on locality, or on
reality in the Einstein-Podolsky-Rosen sense. Thus, our assumption is the
minimal one that is expected to be satisfied by any theory consistent with the
fundamental symmetries like rotational invariance, with or without hidden
variables. Therefore the results are valid for a wide class of theories
including local hidden variable theories, non-local theories, and quantum mechanics.

The proof is generally valid for theories satisfying following criteria.

\begin{enumerate}
\item The theory deals with observations on two particles $A$ and $B$ that
show discrete outcomes, for a particular measurement like the projection of
the angular momentum. Then the product of observed values in one joint
measurement for any relative settings of the apparatus are quantized and
ranges from $+S^{2}$ to $-S^{2}.$ This is of course due to the fact that each
subsystem can take only values from $+S$ to $-S,$ in steps of $1.$ Thus the
theory is not dealing with standard classical measurements where any value
between $+S$ to $-S$ is allowed. For the special case of two-valued outcomes,
$A=\pm1,\quad B=\pm1,$ in units is $S.$ This is the case considered by Bell.

\item The theory of correlations obeys the conservation of angular momentum on
the average over the ensemble, and for the case of singlet state,
$S_{Total}=0,$ there is rotational invariance. Note that this is a weak
assumption, since we do not insist on the validity of the conservation law for
individual events.
\end{enumerate}

The correlation function allowed by the basic assumption of validity of
conservation law is unique, and surprisingly it is identical to the quantum
mechanical correlation function. Therefore, a physical system with discrete
observable values can show correlations different from what is predicted by
quantum mechanics only by violating a fundamental conservation law! It is
possible to extend this analysis to continuous variables, though here we focus
on the experimentally favoured physical systems with discrete outcomes, as
originally considered by Bell. The second criterion is the main assumption,
physically well motivated, in the proof that follows. Since the main
assumption is applied only for ensemble averages and not for individual
events, I do not make any explicit assumption on locality or reality.

To show the inseparable relation between the validity of the conservation law
on the average and the correlation functions in the most transparent way,
first I will derive the correlation function for the spin-1/2 two-particle
system whose total angular momentum is zero. In quantum mechanics this
corresponds to the two-particle singlet state. We will then derive the general
result for higher spin singlet states.

It is \ well known that rotational invariance of the singlet state already
demands that the correlation function is proportional to $\mathbf{a}%
\cdot\mathbf{b,}$ though not enough attention has been paid to the consequence
that any other correlation function will violate the basic symmetry and
conservation law.

Consider a set of measurements in which the direction at $A$ is fixed as
$\mathbf{a}$ and that at $B$ is fixed as $\mathbf{b}$. Each individual
measurement in any direction gives either $+1$ or $-1$ (in unit of $\hbar/2$).
Prepare two subsets of results on the location $A$; one subset containing only
$+1$ and the other containing only $-1.$ For the initial two-particle system
with total zero angular momentum, there will be equal number of $+1$ and $-1$
at $A$ and $B$. The average angular momentum in each subset is given by the
average of the individual values. For the first subset at $A$ this is simply
$+1$ and for the second subset, it is $-1$ (both in units of $\hbar/2$). There
will be two correlated subsets at $B$, whose individual averages will depend
on the setting at $B$. For this one has to use the temporal information for
deciding which measurement at $B$ needs to be correlated with a particular
measurement at $A$. (The situation is symmetric -- we can start with two
subsets of $B$, and then examine the correlated subsets of $A$). If both
analyzers were set for the same direction, $\mathbf{a=b,}$ the (anti)
correlation is perfect \emph{according to the conservation of angular
momentum}, and the average angular momentum at $B$ would be $-1.$ For an
arbitrary direction $\mathbf{b}$ at $B,$ conservation of angular momentum then
predicts that the \emph{average} of the angular momentum measured for the
second particle, correlated with the $+\hbar/2$ values of the first particle,
should be simply the vectorial projection of $-\hbar/2$ onto the direction
$\mathbf{b}$, or just $-\cos(\theta)$ in unit of $\hbar/2,$ where $\theta$ is
the angle between the measurement directions (see figure 1).

It is perhaps important to stress this obvious point further, with a proof, to
preempt a possible confusion. For individual measurements of the two-point
correlation, the conservation law cannot be invoked, since only the
conditional probabilities are predicted by quantum mechanics. Therefore,
measurement of $+1$ on one particle still leaves the outcome on the other
random, with only the probability fixed by the theory. The situation for
averages is very different. If the average spin projection for the $+$subset
on the particles at $A$ gives $S_{A}=+1,$ and if the average at $B$ is
$S_{B}(\theta)$ for the setting of the analyzer at an angle $\theta,$ the
component of average angular momentum at $A$ in the direction identical to the
setting at $B$ is%
\begin{equation}
S_{A,\theta}=S_{A}\cos(\theta)=\cos(\theta)
\end{equation}
Conservation of angular momentum implies that the total angular momentum in
any direction is zero since the initial angular momentum is zero. Therefore
\begin{equation}
S_{tot}=S_{A,\theta}+S_{B}(\theta)=\cos(\theta)+S_{B}(\theta)=0
\end{equation}
This is independent of the theory, and necessarily implies that for any theory
and for experimental data that respect the conservation law, $S_{B}%
(\theta)=-\cos(\theta),$ given that average $S_{A}=+1.$ What is important is
to note that this statement uses only the vectorial properties of the
mechanical quantity, the average angular momentum, and does not involve
counter-factual arguments as it would for individual measurements. Therefore,
irrespective of the theoretical description, angular momentum conservation
will be violated if $S_{B}(\theta)\neq-\cos(\theta)$ when $S_{A}=+1.$%

\begin{figure}
[ptb]
\begin{center}
\includegraphics[
height=2.0998in,
width=3.8917in
]%
{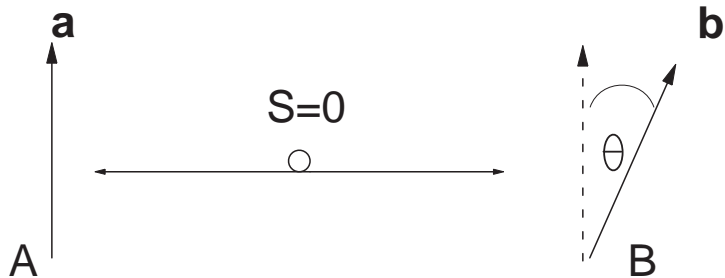}%
\caption{Diagram explaining the settings of the anlayzers. Since the
\emph{average} angular momentum is a vectorial quantity, the average
correlation at angle $\theta$ is $-\cos(\theta),$ when the source has zero
angular momentum.}%
\end{center}
\end{figure}

Similarly, for the subset with results $-1$ for $A$, the average of the
correlated events at $B$ will be the projection of the angular momentum $+1$,
or $+\cos(\theta).$ The \emph{correlation} of the angular momentum between $A$
and $B$ in both cases is the average of the corresponding products in the two
subsets, $[-1_{A}\times\cos(\theta)_{B}+1_{B}\times-\cos(\theta)_{A}]/2,$
which is $-\cos(\theta).$ Therefore the correlation for the entire set is also
$-\cos(\theta)$ or $P(\mathbf{a},\mathbf{b})=-\mathbf{a}\cdot\mathbf{b.}$

It is worth listing the steps clearly leading to the theory-independent
correlation function that follows from the conservation law. The projection of
the classical angular momentum vector in a direction that is rotated by angle
$\theta$ is just $\cos(\theta).$ In a situation in which individual
measurements give discrete values $+1$ and $-1,$ the \emph{average angular
momentum still preserves this relation}. A set of particles with angular
momentum is $+1$ in a particular direction will show an average angular
momentum $\cos(\theta)$ in a direction rotated by $\theta.$

The correlation function of a set of measurements on two particles is
\begin{equation}
P(\mathbf{a},\mathbf{b})=\frac{1}{N}\sum(A_{i}B_{i})
\end{equation}
We consider the situation when the initial angular momentum of the two
particles together is zero. Since $A_{i}$ and $B_{i}$ take values of $+1$ or
$-1$, we group the values such that all $A_{i}$ are $+1$ in the first group
and all $A_{i}$ are $-1$ in the second. Of course the $B_{i}$ are mixed in
both groups. Then the summation index, which is scrambled now due to
regrouping, is relabelled from $1$ to $N$ again, preserving the relative order
and thus the correlation. We \ get
\begin{equation}
P(\mathbf{a},\mathbf{b})_{C}=\frac{1}{N}\sum(A_{i}B_{i})=\left\langle \left(
\frac{1}{N_{A+}}\sum_{j=1}^{N/2}+1.(B_{j})+\frac{1}{N_{A-}}\sum_{j=(N/2)+1}%
^{N}-1.(B_{j})\right)  \right\rangle
\end{equation}
We have used the subscript $C$ to indicate that this correlation function
assumes the validity of the conservation law. The first term is the
conditional average of $B_{j}$ given $A_{j}=1,$ and the second term is the
average of $B_{j}$ given $A_{j}=-1.$ $N_{A+}=N_{A-}=N/2.$ Then we get, using
only the conservation law for the averages,
\begin{equation}
P(\mathbf{a},\mathbf{b})_{C}=\frac{1}{2}\left(  -\cos(\theta\right)
-\cos(\theta))=-\cos(\theta)
\end{equation}
This is same as the quantum mechanical correlation function $P(\mathbf{a}%
,\mathbf{b})_{QM}.$ We have proved that the correlation function $-\cos
(\theta)$ and thus $P(\mathbf{a},\mathbf{b})_{QM}$ is a consequence of the
conservation of angular momentum. The correlation function represents the
conservation law, and follows uniquely and directly from it.

The combination%

\begin{equation}
M_{C}=\left|  P(\mathbf{a},\mathbf{b})-P(\mathbf{a},\mathbf{b}^{\prime
})\right|  +\left|  P(\mathbf{a}^{\prime},\mathbf{b}^{\prime})+P(\mathbf{a}%
^{\prime},\mathbf{b})\right|
\end{equation}
for the theory of correlations satisfying the fundamental conservation law is
thus identical to $M_{QM}$ since $P(\mathbf{a},\mathbf{b})_{C}=P(\mathbf{a}%
,\mathbf{b})_{QM}.$ It follows that $M_{C}$ can exceed $M_{LHVT}$ for several
combinations $\left\{  \mathbf{a},\mathbf{a}^{\prime},\mathbf{b}%
,\mathbf{b}^{\prime}\right\}  .$ \ 

This derivation of the unique correlation function for the spin-1/2
two-particle system from the conservation of angular momentum immediately
implies that any other correlation function with a dependence on the relative
angle other than $-\cos(\theta)$ violates the law of conservation of angular
momentum. Since the Bell's inequalities can be obeyed only when the
correlation function is different from the quantum mechanical correlation
$-\mathbf{a}\cdot\mathbf{b,}$ it follows that a local hidden variable theory
does not respect the law of conservation of angular momentum. In order to get
a deviation from\emph{ }the correlation function $P(\mathbf{a},\mathbf{b}%
)_{C}=-\mathbf{a}\cdot\mathbf{b=}$ $P(\mathbf{a},\mathbf{b})_{QM},$ \emph{the
conservation law for angular momentum has to be violated}. This completes the
proof that a general theory of correlations of discrete two-valued variables
will violate the Bell's inequalities if the theory respects the fundamental
conservation law of angular momentum on the average.

The generalization to higher spins is straightforward. For a spin-$S$
entangled singlet state, there will be $2S+1$ different possibilities for
measurement results, $+S,+(S-1)..0..-(S-1),-S.$ All of these occur with equal
probability, due to rotational invariance, and we group the measurements at
$A$ as earlier in this sequence. For a specific group, say with $(S-n)$ as the
projection, conservation of angular momentum implies that the average of the
angular momentum at $B$ is just the vector projection of the opposite angular
momentum, $-(S-n)\cos(\theta).$ Then the correlation for that group is
\begin{equation}
P(S,\theta)=-(S-n)^{2}\cos(\theta)
\end{equation}
The same correlation will be observed also for the results $-(S-n)$ at $A.$
For the `$0$' state, the correlation (average of the angular momentum
correlation) is zero. Therefore, the average of all the two-point measurements
is%
\begin{align}
P(\mathbf{a},\mathbf{b)}_{C}  &  \mathbf{=}\frac{-\cos(\theta)\left(
S^{2}+(S-1)^{2}+...0+...(S-1)^{2}+S^{2}\right)  }{2S+1}\nonumber\\
&  =-2\cos(\theta)\left(  S^{2}+(S-1)^{2}+...0\right)  /2S+1\nonumber\\
&  =-2\cos(\theta)S(S+1)(2S+1)/6(2S+1)\nonumber\\
&  =-\cos(\theta)S(S+1)/3
\end{align}
This is exactly the quantum mechanical correlation function $P(\mathbf{a}%
,\mathbf{b)}_{QM}$ for the spin-$S$ singlet state \cite{mermin}. Thus, the
physically valid correlation function can be uniquely derived from the
conservation law, and any other correlation function is incompatible with the
requirement of conservation of angular momentum on the average. Note that this
gives the correct $P(\mathbf{a},\mathbf{b)=}-\cos(\theta)/4$ for the
spin-$1/2$ case, or just $-\cos(\theta)$ when \ the observable values are
taken as $+1$ and $-1$ instead of $+1/2$ and $-1/2.$

To be more explicit and precise about the amount of departure from the
conservation law, let us see what the angular dependence of the Bell
correlation function for local hidden variable theory looks like. This
dependence can be derived using the assumption of locality and observing that
the joint probabilities obey the separability of probabilities in the local
hidden variable theories (This is done and discussed in detail in Bell's
writings; see for example ref. \cite{bell-corr}). If the angle of one of the
analyzers is suddenly changed, the correlation function has to change, and
since there is no information available on the instantaneous setting of the
second spatially separated analyzer, this change is to be composed of
individual changes that depends separably on the individual angular setting of
each analyzer. This shows that the correlation function changes linearly with
angle (Fig. 2) \cite{bell-corr}. If the correlation function obeys the
constraint that for $\theta=0$ the correlation is perfect and for $\theta
=\pi/2$ the correlation is zero for the spin-half case, the function is linear
in the entire range. We see that such a correlation function does not respect
the conservation of angular momentum of the total state on the average. Note that it is not possible to attribute the missing angular momentum to the hidden variables, since the theory is local and perfect conservation is seen for special angular settings. %

\begin{figure}
[ptb]
\begin{center}
\includegraphics[
height=1.8965in,
width=2.3229in
]%
{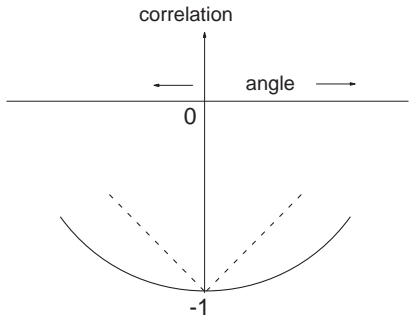}%
\caption{Diagram showing the different dependences of the correlation function
from conservation law (solid curve, same as the QM correlation function) and
the Bell Correlation function, on relative angle. The linear change in
correlation (dotted lines) is a characteristic of local hidden vaiable
theories of the type discussed by Bell. It violates the requirement of
conservation of angular momentum as explained in the text.}%
\end{center}
\end{figure}

The implications of this result to any experiment attempting to test the
Bell's inequalities are obvious. All such experiments are testing for the
possible validity of theories that are incompatible with the fundamental
conservation laws. Given that these conservation laws are consequences of
symmetries, independent of any theoretical formulation, there is no surprise
that experimentally measured correlations follow the expression $P(\mathbf{a}%
,\mathbf{b})_{C}=-\mathbf{a}\cdot\mathbf{b.}$ Any other correlation indicates
a violation of the conservations law on the average. If this result were known before
such experimental attempts started, the attitude towards such tests was likely
to be very different. In any case, it is clear that any expectation of large
deviation from this correlation function, required to obey the Bell's
inequalities, is physically unfounded and in dissonance with the good symmetries.

This result reinforces the statements made earlier in ref. \cite{unni-fpl}
that entanglement in such cases is the general superposition with the
constraint of conservation law of angular momentum. The conservation law is
what is encoded in the relative phase, and in entanglement. Thus the fidelity
with which the conservation law is preserved is directly related to the
measure of entanglement. Processes like decoherence diffuses the individual
phase and thus the relative phase, slowly washing out the fidelity of the
conservation law (preserved however when the total system including the
interacting environment is considered) and hence the entanglement fidelity.
With this insight, it is much easier to understand the subtleties of quantum
entanglement. Applications of this insight will be discussed elsewhere. 

In summary, I have shown that a unique correlation function is associated with
a general theory of correlations of discrete two-valued variables if the
theory respects the fundamental conservation law of angular momentum on the
average. This correlation function turns out to be identical to the quantum
mechanical correlation function for the same experimental situation. I have
also shown that such a correlation function and therefore theories of
correlations that respect the conservation law violate the Bell's
inequalities. Measured correlations of spin projections can obey the Bell's
inequality only by violating the conservation law of angular momentum. A
theory of correlations satisfying Einstein locality and EPR reality satisfies
the Bell's inequality and a theory that satisfies the fundamental conservation
law violates the inequality. Therefore, a theory of correlations satisfying
Einstein locality, reality in the Einstein-Bell sense, and the validity of the
fundamental conservation law cannot be constructed. Some discussion on the
mutual compatibility of the conservation laws and Einstein locality can be
found in ref. \cite{unni-curr,unni-fpl}. A generalization of these results to
continuous observables is in progress.

\bigskip

\noindent Acknowledgements: I thank Matt Walhout for a crucial spontaneous
discussion, and Vikarm Soni for initiating thoughts on the relation between
conservation laws and correlation functions. I thank B. d'Espagnat for a
patient discussion on several related issues. These results were first
presented during the congress of quantum physics at the Centre for Philosophy
and Foundations of Science, Delhi in January 2004.


\begin{thebibliography}{9}                                                                                                %

\bibitem {belin}F. J. Belinfante, \textit{A survey of hidden variable
theories}, (Pergamon Press, Oxford, 1973).

\bibitem {bell}J. S. Bell, \textit{Physics} \textbf{1} (1965) 195 ;
\textit{Speakable and unspeakable in quantum mechanics} (Cambridge University
Press, 1987).

\bibitem {exp}See for a review, \textit{The Einstein, Podolsky, and Rosen
Paradox in Atomic, Nuclear and Particle Physics}, A. Afriat and F. Selleri,
(Plenum Press, New York and London, 1999).

\bibitem {mermin}A. Garg and N. D. Mermin, Phys. Rev. Lett., \textbf{49}
(1982) 901.

\bibitem {bell-corr}J. S. Bell, Einstein-Podolsky-Rosen experiments (Article
10) in \textit{Speakable and unspeakable in quantum mechanics, } (Cambridge
University Press, 1987).

\bibitem {unni-curr}C. S. Unnikrishnan, Current Science \textbf{79} (2000) 195.

\bibitem {unni-fpl}C. S. Unnikrishnan, Found. Phys. Lett. \textbf{15} (2002)
1; available free at present at the Foundations of Physics Letters journal web
site, listed as the most viewed paper (www.kluweronline.com/issn/0894-9875).
\end{thebibliography}
\end{document}